\documentclass[aps,pre,twocolumn,superscriptaddress,letterpaper]{revtex4}

\usepackage{t1enc}
\usepackage{amsmath}
\usepackage{graphicx}
\usepackage{epsfig}
\usepackage{psfrag}
\usepackage{times}

\begin{document}

\title{The frustrated spherical model: an alternative to Ginzburg-Landau Hamiltonians with competing interactions}

\author{Alejandro Mendoza-Coto}
\affiliation{Department of Theoretical Physics, Physics Faculty, University of
Havana, La Habana, CP 10400, Cuba}
\affiliation{``Henri-Poincar\'e-Group'' of Complex Systems, Physics Faculty, University of
Havana, La Habana, CP 10400, Cuba}

\author{Rogelio D\'\i az-M\'endez}
\affiliation{Nanophysics Group, Department of Physics,
Electric Engineering Faculty, 
CUJAE, ave 114 final, La Habana, Cuba}
\affiliation{``Henri-Poincar\'e-Group'' of Complex Systems, Physics Faculty, University of
Havana, La Habana, CP 10400, Cuba}

\date{December 2010}

\begin{abstract}
We solve analytically the Langevin dynamics of the classic spherical model considering the
ferromagnetic exchange and a long-range antiferromagnetic interaction.  Our results in the
asymptotic regime, shows an equivalence in the functionality of the spatial and
self-correlations between this model and the recently studied Ginzburg-Landau frustrated
model within the Hartree approximation. A careful discussion is done about the low
temperature behavior in the context of glassy dynamics.
The appearance of interesting
features regarding the establishment of the ferromagnetic phase is also analyzed in view of
the effects of the spherical restriction. We propose a new variant of the spherical model in
which the global restriction is substituted by an infinite set of restrictions over finite
size regions. This modification leads to a new dynamical equation that suggests the
appearance of the low temperature phase transition even in the non-frustrated case were the
classic spherical model fails.
\end{abstract}

\maketitle

\section{Introduction}

The study of the fundamental role of microscopic interactions on the relevant properties of
magnetic materials involves two main approaches: discrete models, like the Ising model which
is commonly used to represent systems with uniaxial anisotropy, and continuous models among
which the Ginzburg-Landau model is one of the most widely known. These two approaches have
been extensively used both numerically\cite{glei03,Ja2004,lucas07,rdm10} and
analytically\cite{garel82,roland90,pig07,mulet07} to describe equilibrium and dynamical
features observed in real systems.\cite{rose83,seul92,SeAn1995,suizos} Concerning to the
local magnetization bound, unlike the Ising model in which the site magnetization take values
$\pm1$, the Ginzburg-Landau double well potential is the responsible to guaranty  the
non-divergence of the local magnetization. Unfortunately, this potential introduce a strong
non-linearity in the dynamical equations describing  the time evolution of the local order
parameter, and do not allow the magnetization of the system to saturate at high enough
external fields. On the other hand, for discrete models some dynamical treatments are
forbidden due to the natural impossibility of define temporal derivatives of the local
magnetization.

In this context, the spherical model\cite{berlin52,baxter82} is something in between. It is a
linear model that can be treated analytically, while the saturation of the magnetization is
guarantied by means of the spherical constraint. Despite this niceties, this model fails in
obtaining of the ferromagnetic transition in two dimensions. On the basis of this
disagreement it has been argued that the spherical constrain allows the system to be in a
large number of disordered configurations. In two dimensions this degeneracy prevents the
system to reach a ferromagnetic phase transition at any non-zero temperature. To manage this
difficulty a lot of work have been carried out making some radical modifications as, for
example, the substitution of the short range exchange interactions by a long range
ferromagnetic one.\cite{brody94,cannas00}

Recently, a great effort has been done in the study of the system with competing isotropic
interactions, i. e. systems whose fluctuation spectrum has a minimum in a non-zero radius
spherical shell in the inverse space.\cite{grousson00,grousson02,pro08,tar07,nm10} In
magnetic systems, these interaction are usually represented by the ferromagnetic exchange
interaction and the anti-ferromagnetic long-ranged dipolar interaction. The Langevin dynamic
of a Ginzburg-Landau Hamiltonian with such interactions have been studied previously by means
of the Hartree self-consistent field approximation.\cite{mulet07,nm10} In these works the
authors confirm the theoretical picture found analytically by means of numerical simulations.
Moreover, the symmetry breaking properties of the model were also studied by means of a very
similar procedure.\cite{nmprl} In particular, it was shown that taking the infinite time
limit it was possible to encounter the phase transition to the modulated ordered state by
means of purely dynamical calculations.

In the present work we solve the Langevin dynamics of the spherical model considering the
common square field-gradient attractive term plus a repulsive long-range antiferromagnetic
term in the long times limit. We show that, starting from a disordered initial condition,
quenches to the $T=0$ regime develops a long time dynamics very close to that encountered in
Ref. \cite{nm10}. As it is expected, for this frustrated spherical model, the value of the
critical field dividing the two different dynamical regimes is also the saturation value. For
$T>0$ we found a standard paramagnetic relaxation after certain characteristic time.
Anyway, the divergence of this time for low temperatures as well as the existence of
modulated domains of typical length can be related to the reminiscence
of a glassy region not completely verified for our model. The monotony of the $h$-$T$ curve
for a fixed magnetization is also an interesting finding that connects the results with
recent experimental works.
Finally, we propose a new variant of the
spherical model in which the global restriction is substituted by an infinite set of
restrictions over finite spatial regions of the system. This lead to a new dynamical equation
that is also presented in detail. We discuss the way in which this modification suggest the
appearance of the low temperature phase transition even in the non-frustrated case were the
spherical model fails.

The rest of the paper is organized as follows. The next section is completely devoted to the
presentation of the formalism and the solution of the dynamical equation for all the regimes
involving the value of the applied external field. Some commonly studied observables as
spatial and self-correlations are presented and discussed in some detail in section \ref{ob}
comparing its asymptotic expressions with those recently obtained for the Ginzburg-Landau
model. In section \ref{prc} the more relevant findings are discussed in the light of the
known phenomenology of related models and some experimental observations, in order to better
understand the advantages of using the spherical model in the context of competitive
interactions. Also a new dynamical equation is obtained by means of suitable modifications to
the standard model and is discussed the way in which this new approach may lead to the
appearance of the phase transition. We summarize the most important results and discussions
in section \ref{conc}. In appendix \ref{ap1} the spatial correlations of some particular
configurations are calculated by means of geometrical considerations.

\section{Model and Formalism}
\label{mf}

We consider the following Hamiltonian dependent on the local parameter $\phi(\vec{x},t)$
\begin{eqnarray}
\nonumber
{\cal H}[\phi(\vec{x})]&=&\frac{1}{2}\int_\Omega d^dx\left[(\vec{\nabla}\phi)^2-2 h \phi(\vec{x})\right]
\\
&-&\frac{1}{2\delta}\int_\Omega\int_\Omega d^dx\ d^dx' J(\vec{x}-\vec{x}')\ \phi(\vec{x})\ \phi(\vec{x}')
\end{eqnarray}
where $J(\vec{x}-\vec{x}')$ represents a repulsive, isotropic, competing interaction and
$\Omega$ is the $d$-dimensional volume of the system, which is finally taken as infinity. The
homogeneous external field is represented by $h$ and the parameter $\delta$ measures the
relative intensity between the
 attractive and repulsive interactions.
The local field is subjected to the restriction
\begin{equation}
\int_\Omega \phi^2(\vec{x})\ d^dx = m^2 \Omega
\label{rest}
\end{equation}
that represents a system with local magnetization $m$. In order to include this restriction
in the dynamics it is widely used the effective Hamiltonian\cite{let95}
\begin{equation}
{\cal H}'[\phi(\vec{x})]={\cal H}[\phi(\vec{x})]+\frac{\lambda(t)}{2}\left[\int_\Omega \phi^2(\vec{x}) d^dx -m^2\Omega\right]
\end{equation}
and in consequence the Langevin equation of motion is given by
\begin{eqnarray}
\nonumber
\frac{\partial \phi(\vec{x},t)}{\partial t}&=&\nabla^2\phi(\vec{x})-
\frac{1}{\delta}\int_\Omega d^dx' J(\vec{x}-\vec{x}') \phi(\vec{x}')\\
&+&h-\lambda(t)\phi(\vec{x})+\eta(\vec{x},t)
\label{eqm}
\end{eqnarray}
the solution of this equation is carried out in the inverse space, defined by the following Fourier
transforms
\begin{eqnarray}
\nonumber
\phi(\vec{x})&=&\int\frac{d^dk}{(2\pi)^d} e^{i \vec{k}\cdot\vec{x}} \hat{\phi}(\vec{k})\\
\hat{\phi}(\vec{k})&=&\int d^dx\ e^{-i \vec{k}\cdot\vec{x}} \phi(\vec{x})
\end{eqnarray}
The equation of motion in the momenta space is given by
\begin{eqnarray}
\nonumber
\frac{\partial\phi(\vec{k},t)}{\partial t}&=&-\left[k^2+\frac{1}{\delta} J(k)+ \lambda(t) \right] \phi(\vec{k},t)\\
&+&h(\vec{k})+\eta(\vec{k},t)
\end{eqnarray}
and its solution can be written in terms of the response function $R(\vec{k},t,t')$ as
\begin{eqnarray}
\nonumber
\phi(\vec{k},t)&=&\phi(\vec{k},0) R(\vec{k},t,0)+\int_0^t R(\vec{k},t,t')\eta(\vec{k},t') dt'\\
&+&\int_0^t R(\vec{k},t,t') h(\vec{k}) dt'
\label{phik}
\end{eqnarray}
where
\begin{eqnarray}
R(\vec{k},t,t')=\frac{Y(t')}{Y(t)} e^{-\hat{A}(k)(t-t')}
\end{eqnarray}
It has been also defined $\hat{A}(k)=A(k)-A(k_0)$, $Y(t)=e^{\int_0^t I(t') dt'}$,
$I(t)=A(k_0)+\lambda(t)$ and $A(k)=k^2+\frac{1}{\delta}J(k)$. At this point the problem has
been formally solved, except for the fact that the parameter $\lambda(t)$ remains unknown. It
can be shown that the spherical restriction leads to the normalization condition
$C(t,t)=m^2$, where $C(t,t')$ represents the self-correlation function, obtained from the
usual correlation function given by
\begin{equation}
C(\vec{k},\vec{k}',t,t')=\langle\phi(\vec{k},t)\phi(\vec{k}',t')\rangle
\end{equation}
Thus, to  impose the normalization condition for the self-correlation function allows to
obtain the following equation
\begin{eqnarray}
\nonumber
m^2K(t)&=&\Delta  f(t) + 2 T \int_0^t K(\tau) f(t-\tau) d\tau \\
&+& h^2 \left[\int_0^t K^{1/2}(\tau) e^{-\hat{A}(0)(t-\tau)} d\tau \right]^2
\label{eqk}
\end{eqnarray}
where $K(t)=Y(t)^{\frac{1}{2}}$ and
\begin{equation}
f(t)=\frac{1}{(2\pi)^d} \int d^dk\ e^{-2\hat{A}(k)t}
\end{equation}
In systems with isotropic competing interactions the fluctuation spectrum $\hat{A}(k)$ has a
minimum in a shell in the inverse space of certain radius $k_0$. As it was shown in previous
works\cite{mulet07,nm10} the long time behavior of $f(t)$ is dominated by the local topology
of the fluctuation spectrum around $k_0$. In consequence, it is enough to consider an
expansion of the fluctuation spectrum $\hat{A}(k)$ up to second order around its minimum,
such expansion has the form
\begin{equation}
\hat{A}(k)=\frac{A_2}{2}(k-k_0)^2
\label{hAk}
\end{equation}
where $A_2=\frac{d^2\hat{A}(k)}{dk^2}$. With these ingredients it  is straight forward to
obtain the following asymptotic result
\begin{equation}
f(t)=\frac{C}{t^{1/2}}
\label{ft}
\end{equation}
where $C$ is certain uninteresting constant depending on system dimension. At this point we
carry out the solution of the problem separately, first we analyze the zero temperature case
and then the non-zero temperature case.

\subsection{Case $T=0$}
In order to solve equation (\ref{eqk}) at zero temperature, we found suitable to define the function
\begin{equation}
S(t)=\int_0^t K^{1/2}(\tau) e^{-\hat{A}(0)(t-\tau)} d\tau
\end{equation}
and then to split up the original equation into the following set of equations
\begin{eqnarray}
\nonumber
m^2 K(t)&=&\Delta f(t)+h^2 S^2(t)\\
\frac{dS(t)}{dt}&=&-\hat{A}(0)S(t)+K(t)^{\frac{1}{2}}
\end{eqnarray}
with the initial conditions $K(0)=1$ and $S(0)=0$. Note that this system can be put in the
form
\begin{equation}
\frac{dS(t)}{dt}=-\hat{A}(0)S(t)+\frac{1}{m}\left(\Delta f(t)+h^2S^2(t) \right)^{\frac{1}{2}}
\label{S}
\end{equation}
So, we have to find the asymptotic solution of the above differential equation. In order to
do this we compare the terms $f(t)$ and $S^2(t)$. In general we can see that there are
several possibilities varying $h$. For small  fields $f(t)$ dominates the dynamics and in
consequence $S^2(t)\propto f(t)$. On the other hand, fields higher than certain critical
value $h_c$ leads to a different behavior. In what follows we obtain the different solutions
for $S(t)$ and $K(t)$ varying the external field $h$.

\subsubsection{$h<h_c$}
In the case of small enough fields we can obtain that $S^2(t)=\lambda f(t)$. This
consideration implies the relation
\begin{equation}
\frac{1}{m}\left(\Delta f(t)+h^2S^2(t) \right)^{\frac{1}{2}}=\frac{1}{m}\left(\Delta+h^2\lambda\right)^{\frac{1}{2}}f(t)^{\frac{1}{2}}
\end{equation}
This leads to the simplified equation
\begin{equation}
\frac{dS(t)}{dt}=-\hat{A}(0)S(t)+\frac{1}{m}\left(\Delta+h^2\lambda\right)^{\frac{1}{2}}f(t)^{\frac{1}{2}}
\end{equation}
whose asymptotic solution is given by
\begin{equation}
S(t)=\frac{1}{m\hat{A}(0)}\left(\Delta+h^2\lambda\right)^{\frac{1}{2}}f(t)^{\frac{1}{2}}
\end{equation}
At this point to consider the proposed ansatz implies that
\begin{equation}
\lambda=\frac{\Delta}{m^2\hat{A}(0)^2-h^2}
\end{equation}
and
\begin{equation}
K(t)=\lambda\hat{A}(0)^2f(t)
\end{equation}
It is worth to note that from the definition of $\lambda$ one can see the existence of a
critical field $h_c=m\hat{A}(0)$.

\subsubsection{$h=h_c$}
We start by writing equation (\ref{S}) in the form
\begin{equation}
\frac{dS(t)}{dt}=-\hat{A}(0)S(t)+\frac{hS(t)}{m}\left(1+\frac{\Delta f(t)}{h^2S^2(t)}\right)^{\frac{1}{2}}
\end{equation}
In the long time limit $\Delta f(t)\ll h^2S^2(t)$, then it is possible to write
\begin{equation}
\frac{dS(t)}{dt}=-\hat{A}(0)S(t)+\frac{hS(t)}{m}\left(1+\frac{\Delta f(t)}{2h^2S^2(t)}\right)
\end{equation}
considering an expansion up to first order of the non-linear contribution. Note that if we
take
 now $h=h_c=m\hat{A}(0)$, the first order contribution of the non-linear term cancels and we have
\begin{equation}
\frac{dS(t)}{dt}=\frac{\Delta}{m^2\hat{A}(0)}\frac{f(t)}{2S(t)}
\end{equation}
whose solution yields
\begin{equation}
S^2(t)=\frac{\Delta}{m^2\hat{A}(0)}\int^tf(\tau)d\tau
\label{s2hc}
\end{equation}
and
\begin{equation}
K(t)=\frac{\Delta\hat{A}(0)}{m^2}\int^tf(\tau)d\tau
\label{khc}
\end{equation}
Expressions (\ref{s2hc}) and (\ref{khc}) has no lower integration limit. This is related with
the fact that, in the long time limit, any constant can be neglected in comparison with the
increasing functions $S(t)$ and $K(t)$.

\subsubsection{$h>h_c$}
In this case, from equation (\ref{S}) in the long time regime, it is fairly easy to see that
\begin{equation}
S(t)\propto e^{\left(-\hat{A}(0)+\frac{h}{m}\right)t}
\end{equation}
and in consequence
\begin{equation}
K(t)\propto e^{2\left(-\hat{A}(0)+\frac{h}{m}\right)t} \label{kh>hc}
\end{equation}

\subsection{Case $T>0$}
At finite temperature the equation (\ref{eqk}) contains a convolution term with the function
$f(t)$, which leads always to a exponential long time behavior, as have been obtain
previously in literature. In order to clarify this point, we propose the ansatz\cite{nm10}
\begin{equation}
S(t)=\int_0^t K^{1/2}(\tau) e^{-\hat{A}(0)(t-\tau)} d\tau=\xi K^{\frac{1}{2}}(t)
\end{equation}
for the asymptotic dynamic. By doing this, equation (\ref{eqk}) turns in an integral equation
whose solution can be found by means of the Laplace transform. Using expression (\ref{ft})
for $f(t)$ we get
\begin{equation}
K(t)\propto e^{b^2t}
\end{equation}
where $b$ and $\xi$ are the solutions of the following system of equations
\begin{equation}
b=\frac{2TC\Gamma(\frac{1}{2})}{m^2-h^2\xi^2} \ , \ \ \ \ \ \ \ \ \xi=\frac{2}{b^2+2\hat{A}(0)}
\label{syst}
\end{equation}
Finally, it is worth to note that the above system of equations always give rise to positive
values for both, $b$ and $\xi$. This means that at any finite temperature the model reaches
the same asymptotic dynamics.

\section{Observables}
\label{ob}

At this point we have been able to find the asymptotic behavior of the function $K(t)$ for
any value of temperature and external field. This allows the calculation of some commonly
studied statistical observables. In particular, we obtain in this section the mean value of
the local field $\phi$ (magnetization), the self-correlation function and the spatial
correlation function. For some of these magnitudes closed expressions are found in terms of
the function $K(t)$.

\subsection{Mean value}
\begin{figure}[!htb]
\includegraphics[width=8cm,height=5.7cm]{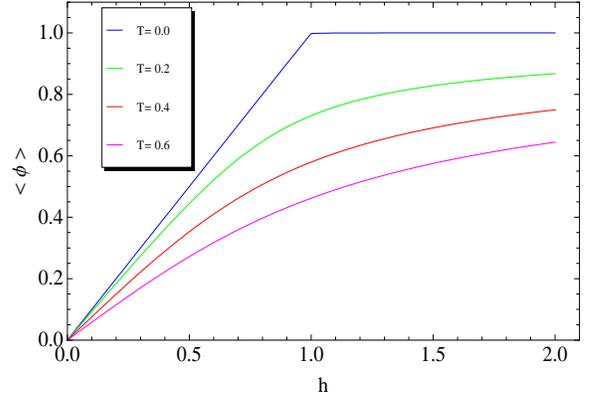}
\caption{Dependence of the mean value of the local field $\langle\phi\rangle$ with the
external field $h$ for several temperatures.} \label{MagVsTH}
\end{figure}

Starting from expression (\ref{phik}) we obtain the following expression for the mean value
\begin{equation}
\left\langle\phi\right\rangle=h \ \xi(h,T)
\end{equation}
where
\begin{equation}
\xi(h,T)=\lim_{t\rightarrow\infty}\frac{S(t)}{K(t)^{\frac{1}{2}}}
\end{equation}
This expression implies that at $T=0$
\begin{equation}
\langle\phi\rangle=\left\{ \begin{array}{ll}
\frac{h}{\hat{A}(0)}                 &  \textrm{for $h<h_c$} \\ \\
 m                                &  \textrm{for $h\geq h_c$} \\
\end{array}
\right.
\label{aaa}
\end{equation}
while for any finite temperature $\xi(h,T)$ is given by the solution of the system (\ref{syst}). In order to
clarify such a dependency of the mean value with the external field and temperature we define the following
non-dimensional magnitudes
\begin{eqnarray}
\frac{h}{h_c}\rightarrow {h}, \ \ \ \ \frac{C\Gamma(\frac{1}{2})T}{\hat{A}(0)^{\frac{1}{2}}m^2}\rightarrow {T}, \ \ \ \ 
\frac{\left\langle\phi\right\rangle}{m}\rightarrow{\phi}
\end{eqnarray}
which are the variables uses in the figure \ref{MagVsTH}.

As can be seen from figure \ref{MagVsTH}, at zero temperature and $h<h_c$ the system behaves
like a purely paramagnetic material with a completely linear relation between magnetization
and external field. On the other hand, for any
 finite temperature the relation between the mean value and the external field is no longer linear
in such a way that $\langle\phi\rangle$ is always an increasing function of the applied
external field.

\subsection{Self-correlation function}
In order to look inside the ordering process we calculate now the self-correlation function,
defined as
$C(t,t')=\left\langle\phi(\vec{r},t)\phi(\vec{r},t')\right\rangle-\left\langle\phi(\vec{r},t)\right\rangle
\left\langle\phi(\vec{r},t')\right\rangle$. In terms of the function $K(t)$ this magnitude
can be written
\begin{eqnarray}
\nonumber
C(t,t')&=&\frac{\Delta}{\sqrt{K(t)K(t')}}f\left( \frac{t+t'}{2}\right) \\ \nonumber
&+&\frac{2T}{\sqrt{K(t)K(t')}}\int_0^{t'}K(\tau)f\left( \frac{t+t'}{2}-\tau\right)d\tau \\ \nonumber
\end{eqnarray}
In this way at $T=0$ we obtain
\begin{equation}
C(t,t')\propto\left\{ \begin{array}{ll}
\frac{(4tt')^{\frac{1}{4}}}{(t+t')^{\frac{1}{2}}}                 &  \textrm{for $h<h_c$} \\ \\
\frac{1}{\left(t+t'\right)^{\frac{1}{2}}(4tt')^{\frac{1}{4}}}                                  &  \textrm{for $h=h_c$} \\  \\
\frac{e^{-(\hat{A}(0)+\frac{h}{m})(t+t')}}{\left(t+t'\right)^{\frac{1}{2}}}      &  \textrm{for $h>h_c$} \\
\end{array} \right.
\end{equation}

Considering our definition of $C(t,t')$, it can be seen that for high enough fields ($h\geq
h_c$) the system presents an interrupted aging and eventually evolves to a ferromagnetic
configuration in which there are no fluctuations at all.  On the other hand, for $h<h_c$ the
system shows a slow relaxation typical of a coarsening scenario.

For $T>0$ the long time behavior $(b^2t'\gg1)$ of the self-correlations is given by
\begin{equation}
C(t,t')=2T\int_{\frac{t-t'}{2}}^{\frac{t+t'}{2}}f(\tau)e^{-b^2\tau}d\tau
\label{sco}
\end{equation}
showing an interrupted aging and an asymptotic temporal translational invariant relaxation
$C(t,t')=C(t-t')$ typical of systems in thermal equilibrium. This means that for any finite
temperature the system always reach the equilibrium state.

\subsection{Spatial correlation function}
With the aim of study the texture of the system under different conditions, we calculate the
spatial correlation function. In our case the isotropy  of interactions and initial
conditions ensures an homogeneous mean value of the order parameter. The spatial correlation
function is defined as
$C(R,t)=\langle\phi(\vec{x},t)\phi(\vec{x}+\vec{R},t)\rangle-\langle\phi(\vec{x},t)
\rangle\langle\phi(\vec{x}+\vec{R},t)\rangle$. At zero temperature ($T=0$) we obtain
\begin{equation}
C(R,t)=\frac{\Delta}{K(t)}\int \frac{d^d\vec{k}}{(2\pi)^d}e^{-2\hat{A}(\vec{k})t}e^{i\vec{k}\cdot\vec{R}}
\label{CRt}
\end{equation}
and assuming $k_0R\gg1$ it is possible to obtain
\begin{eqnarray}
\nonumber
C(R,t)&\propto&\frac{f(t)}{K(t)}\frac{\Delta}{(k_0R)^{\frac{d-1}{2}}}\\
&&\mathrm{cos}\left(k_0R-\frac{(d-1)\pi}{4}\right)e^{-\frac{R^2}{4A_2t}} \label{crr}
\end{eqnarray}
where  $k_0$ and $A_2$ are defined by the expression (\ref{hAk}). At zero temperature the
relation $f(t)/K(t)$ contains the dependence with the external field. For $h<h_c$ this
functionality tends to a constant value, while in the other cases we have
\begin{equation}
\frac{f(t)}{K(t)}\propto\left\{ \begin{array}{ll}
\frac{1}{t}                                  &  \textrm{for $h=h_c$} \\  \\
\frac{e^{-2\left(-\hat{A}(0)+\frac{h}{m}\right)t}}{t^{\frac{1}{2}}}      &  \textrm{for $h>h_c$} \\
\end{array} \right.
\end{equation}
It can be noticed that for fields equal or higher than the critical one the system evolves to
the completely ferromagnetic configuration. This can be appreciated in the decay of the
spatial correlation function. On the other hand, for $h<h_c$ the presence of the exponential
term reveals a phenomenology of growing domains with typical length $L(t)\propto
t^{\frac{1}{2}}$, very common in models with non-conserved order parameter. These domains
represent random oriented modulated structures with frequency $k_0$. It is worth to note that
the power law decay do not imply the presence of structural defects necessarily, it is
possible to show how such a term arise from the average over the ensemble of possible
configurations (see Appendix \ref{ap1}).

When $T>0$ it is possible to obtain in the equilibrium state
\begin{eqnarray}
\nonumber C(R)&=&\lim_{t\rightarrow\infty}2T\int
\frac{d^d\vec{k}}{(2\pi)^d}e^{i\vec{k}\cdot\vec{R}}\\ \nonumber
&&\int_0^td\tau \frac{K(\tau)}{K(t)}e^{-2\hat{A}(\vec{k})(t-\tau)}\\
&=&\frac{2T}{(2\pi)^d}\int d^d\vec{k}\ \frac{e^{i\vec{k}\cdot\vec{R}}}{b^2+A_2(k-k_0)^2}
\end{eqnarray}
where the last integral have been calculated before in Ref. \cite{mulet07} and gives
\begin{equation}
C(R)\propto\frac{1}{(k_0R)^{\frac{d-1}{2}}}e^{-\frac{R}{\xi}}\mathrm{cos}\left(k_0R-\psi\right)
\label{crfinita}
\end{equation}
with
\begin{eqnarray}
\nonumber
\xi&=&\frac{\sqrt{A_2}}{b}\\ 
\psi&=&\frac{(d-1)\pi}{4}-\frac{d-1}{2}\mathrm{tan}^{-1}\left(\frac{1}{k_0\xi}\right)
\label{chii}
\end{eqnarray}

Expression (\ref{crfinita}) describes a long times state in which a mosaic of modulated structures
establishes with a wave length $2\pi/k_0$ and a typical dimension equal to the correlation length.
As can be seen from (\ref{chii}) this correlation length diverges as temperature tends to zero for
$h<h_c$, and decreases when the field is augmented. It is worth to note that for $h\geq h_c$ the
spatial correlation as has been defined tends to zero when $T\rightarrow0$ indicating an evolution
to the homogeneous state.

\section{Putting results in context}
\label{prc}

The results exposed above can also be viewed in the frame of the general scheme currently
established in literature. From a static point of view, the infinite time configuration of
the system shows no long range stripes order at any finite temperature. This is consistent
with the isotropy of our initial conditions and the fact that there is no symmetry breaking.
In fact, what we do is to study the long times dynamic of the ensemble of disordered
quenches. Anyway, as can be seen in appendix \ref{ap1}, the lack of long range order in this
case is not due to the ensemble degeneration but to the appearance of randomly oriented
domains of typical size.

On the other hand, it has been reported in literature\cite{tar07,tar07cm} that below certain
temperature the Ginzburg-Landau frustrated system behaves as a typical glass. Even when the
spatial correlation found here at finite temperatures (\ref{crfinita}) is consistent with the
glass configuration, the most basic glassy property, that is, the aging behavior of the
self-correlations, is not verified in the long times limit. Instead, as we have already
pointed, there exists interrupted aging for all non-zero temperature, so the system develops
an aging dynamics up to a finite time $\tau_e$ that diverges only at $T=0$.
\begin{figure}[!htb]
\includegraphics[width=8cm,height=5.7cm]{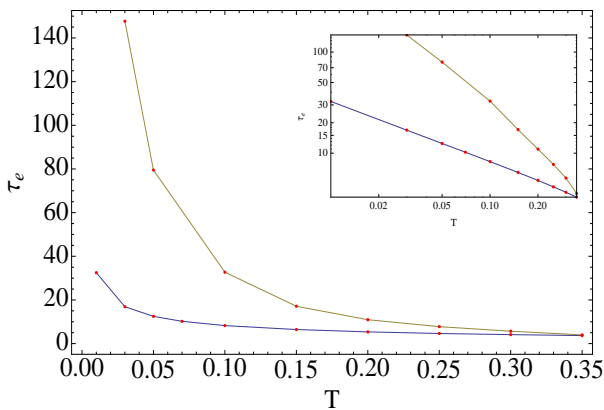}
\caption{Characteristic time $\tau_e$ up to which the aging dynamics can be observed. Upper
curve corresponds to a field $h=0.5 h_c$ and lower curve corresponds to $h=h_c$. The inset
represents  the logarithmic plot of the main figure.} \label{taus}
\end{figure}
This characteristic time can be obtained by fitting the numerical integration of equation
(\ref{eqk}) with the asymptotic exponential behavior and is shown as a function of the
temperature in figure \ref{taus}. We can see from the inset that the characteristic time
behaves as a power law in a vicinity of $T=0$. As it have been suggested,\cite{deAle} this
behavior remains very close to that of the glass in the sense that one can always find a
temperature below which the aging dynamic establishes up to a time far beyond any laboratory
scale.

It is worth to note that the divergence of the correlation length $\xi$ of the ordered
domains at $T=0$ have been reported in a number of glassy systems.\cite{tar07cm} The spatial
correlation (\ref{crr}) certainly accounts for such a glassy configuration. As it is
demonstrated in Appendix \ref{ap1}, this correlation functionality corresponds to an ensemble
of perfect striped configurations every one of which has long range order in a particular
direction. This $T=0$ spatial correlations do not show any more the typical exponential decay
but a potential decay term $1/R^{(d-1)/2}$ that should not be straightforwardly interpreted
as a direct evidence of the presence of topological defects for $d>1$. Correlations obtained
by means of the average over the ensemble of perfect modulated structures (i.e. considering
all the possible  orientations) must be in general a decreasing function of $R$ for $d\geq2$,
even when we have long range order in a particular direction (see Appendix \ref{ap1}). For
the $T=0$ case the glassy behavior is reinforced from the fact that equilibrium is never
reached in any finite time.

Yet, there is another arising feature of this model making it interesting for future studies
in the context of magnetic frustrated systems. If we go back to figure \ref{MagVsTH} it may
be noticed the non-trivial fact that lower temperatures requires lower field values to reach
the same magnetization. In our spherical model approach the completely ferromagnetic phase
can be observed by means of a total saturation $\langle\phi\rangle=1$ reached only at T=0.
These features may be suggesting that, for systems described with this approach, the
ferromagnetic configuration must be reached at low temperatures by applying lower fields.
This interesting result has been recently observed in experimental
systems\cite{suizos,suizos2} and is opposed to the commonly accepted phenomenology for the
Ginzburg-Landau model. \cite{gennes72,garel82,spivak04}

\begin{figure}[!htb]
\includegraphics[width=8cm,height=5.7cm]{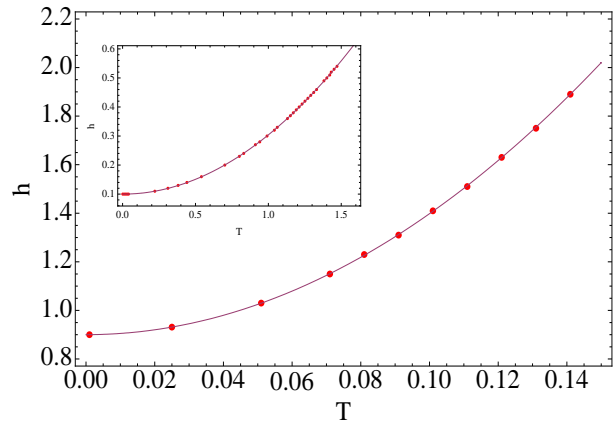}
\caption{External field as a function of temperature for a given value of the magnetization.
Dotted curve in the main figure corresponds to $\langle\phi\rangle=0.9$ while the inset
corresponds to $\langle\phi\rangle=0.1$. The continuous lines are fits to quadratic
functionalities.}
\label{hvsT}
\end{figure}

From numerical data of figure \ref{MagVsTH} it has been constructed figure \ref{hvsT} in such a way
that the plotted curves divides the $h$-$T$ space in two regions.
In the upper(lower) region magnetization is greater(lower) than certain value
$\langle\phi\rangle>\phi_0$ ($\langle\phi\rangle<\phi_0$) with $\phi_0=0.9$ for the main
figure and $\phi_0=0.1$ for the inset. The monotony and functionality of curves in figure
\ref{hvsT} is independent of the specific value of $\phi_0$ even in the vicinity of
$\phi_0=1$ that represents the limit of the modulated structures. In other words, if we
define as ferromagnetic a more real not-completely-saturated state where
$\langle\phi\rangle>1-\epsilon$, then figure \ref{hvsT} would be in disagreement with
theoretical results for other thin film models.\cite{garel82} Since $\epsilon$ can be a
sufficiently small value, the last definition is pretty near to what can be observed
experimentally.

This phenomenology is in good agreement not only with experimental works\cite{suizos2} but
also with some analytical approaches of ultrathin magnetic films considering only striped and
uniform phases.\cite{aba95} However, as can be seen from the figure, we obtain a quadratic
fit for the temperature functionality that is valid for all values of $\phi_0$, instead of
the more complex scaling functionality experimentally observed\cite{suizos} that is dependent on the
critical temperature.

So far, it is not easy to understand where this phenomenology encountered in the frustrated
spherical model begins to be disturbed from the fact that this model does not recover the
phase transition, even in the pure ferromagnetic case in $d=2$. Nevertheless, there is an
acceptable agreement with the standard features established in literature for frustrated
systems and, moreover, the fact that
the local field is bounded in some way seems to have good implications in modeling the behavior
of the system in the presence of an external field at $T>0$. This suggests the necessity  of
finding a near model in which the fundamental limitations already mentioned could be overcome.

\subsection{Extension of the model}
\label{exmo}

Any modification of the standard spherical model must be done in a way that improves the
basic characteristics involved in the probable establishment of the ordered phase. Lets consider
again the spherical restriction (\ref{rest}) but now writing it in the local way
\begin{equation}
 \int_{V_{\vec{n}}}\phi^2(\vec{x})d^dx=m^2V
 \label{mrest}
\end{equation}
where $\vec{n}$ is a discrete $d$-dimensional counter vector and $V_{\vec{n}}$ represents certain
$d$-dimensional volumes of finite constant size. These finite volumes do not overlaps each other and
completely fill the spatial region $\Omega$, so the limit $\Omega\rightarrow\infty$
implies an infinite number of such finite regions. It is over these finite volumes that the spherical
constrain is now performed. This is a closer approximation to what actually happens in many real
system.

Following the same Langevin formalism as in section (\ref{mf}) but now including the multiple
restrictions (\ref{mrest}) we reach to the dynamical equation
\begin{eqnarray}
\nonumber
\frac{\partial \phi(\vec{x},t)}{\partial t}&=&\nabla^2\phi(\vec{x})-
\frac{1}{\delta}\int_\Omega d^dx' J(\vec{x}-\vec{x}') \phi(\vec{x}')\\
&+&h-\lambda_{\vec{n}}(t)\phi(\vec{x})+\eta(\vec{x},t)
\end{eqnarray}
that is virtually similar to that of (\ref{eqm}) with the exception that $\lambda$ is now also
dependent on the space region through the counter $\vec{n}$. It means that there is an infinite set of Lagrange multipliers
each one co\-rres\-pon\-ding to a single finite region $V_{\vec{n}}$.

It is completely plausible to expect that the particular geometry of these finite volumes do not
implies substantial difference in the phenomenology of the model. So we may take them as
$d$-dimensional cubes with a linear size $2a$. In this assumption it is possible to write the
equation of motion in the inverse space as
\begin{eqnarray}
\nonumber
\frac{\partial\phi(\vec{k},t)}{\partial t}&=&-\left[k^2+\frac{1}{\delta} J(k)\right]
\phi(\vec{k},t)+h(\vec{k})+\eta(\vec{k},t)\\
&+& \sum_{\vec{n}}\lambda_{\vec{n}}(t)\int\frac{d^dk'}{(2\pi)^d}\phi(\vec{k}',t)
\hat{Q}_{\vec{n}}(\vec{k}-\vec{k}')
\label{eqkmod}
\end{eqnarray}
where
\begin{eqnarray}
\nonumber
\hat{Q}_{\vec{n}}(\vec{k}-\vec{k}')&=&
\int_{V_{\vec{n}}} e^{i(\vec{k}-\vec{k}')\cdot\vec{x}} d^dx
\\
&=&\left[
\prod_{\alpha=1}^d \frac{2 \sin[(k'_\alpha-k_\alpha)a]}{k'_\alpha-k_\alpha}
\right] e^{i(\vec{k}'-\vec{k})\cdot\vec{r}}
\end{eqnarray}
and $\vec{r}=(2\vec{n}+\vec{1}) a$.

The more complex but yet linear model thus modified
is a very good candidate for observing the transition to the ordered phase
in lower dimensions. The modification is substantial  with respect to the standard spherical
model in which temperature can always destroy
the saturated phase, driving the system to multiple non-homogeneous configurations and yet holding the
spherical constraint. In this new approach infinite constraints guaranties that, below certain temperature,
once the correlation length becomes greater than the typical size $a$ of the spherically
constrained volumes, these volumes behaves as fully-magnetized entities not allowing internal
inhomogeneities. This fact may eventually drive the system to the Ising
universality class or, at least, to a new interesting phenomenology in which the
statistical behavior must have a change around some temperature value dependent on the
physically meaningful parameter $a$.

It is worth to note that equation (\ref{eqkmod}) admits the trivial paramagnetic solution as the
limit case in which all Lagrange multipliers are identical $\lambda_{\vec{r}}=\lambda$. As can
be demonstrated
\begin{equation}
\sum_{\vec{n}} \hat{Q}_{\vec{n}}(\vec{k}-\vec{k}')=(2\pi)^d \delta(\vec{k}-\vec{k}')
\end{equation}
and then equation (\ref{eqkmod}) becomes the already solved equation (\ref{eqk}). This limit is
thus associated in the modified model with the high temperatures region. In this conditions
the correlation length is so small that the spherical restriction over the complete system has
the same effect that if it were done over the finite volumes.

\section{Conclusions}
\label{conc}

The Langevin dynamics of the spherical model has been solved in the long times limit
considering isotropic competing interactions. The general asymptotic functionality of
spatial and self-correlations is found to be equivalent to that of Ginzburg-Landau model
in the Hartree self-consistent field approximation.\cite{nm10}
The work was done studying the ensemble of disordered realizations, and we do not verify
any standard glassy transition as that predicted for near models.\cite{tar07} Nevertheless,
for low temperatures the system evolves to a mosaic of randomly oriented  modulated structures,
very similar to that of the glass state. And moreover, if temperature is sufficiently small,
this configuration develops an extended aging dynamics up to times far beyond any laboratory
scale.

We also found that, for any fixed magnetization,  
the $h$-$T$ curve has a quadratic, growing functionality.
In particular, very near the saturation, where the
real system must pass from the modulated to the ferromagnetic state, the monotony
of this functionality is a very important issue. 
It suggests a possible interpretation based on the spin-bound for
recently experimental observations on the phase diagram of ultrathin magnetic films.\cite{suizos}
Anyway, the fundamental impossibility of the classical spherical model to see the phase
transition is also the responsible of the non-saturation of the system at any finite
temperature. In this sense, a model carrying the nice properties arising from the
spherical restriction, but capable, as the Ginzburg-Landau,  to account for the
phase transition,\cite{nmprl} would be an important step forward.

Such a model is introduced by means of the substitution of the global restriction by an
infinite set of restrictions over finite size regions.
The new dynamical equation is derived, and it is shown that this more complex but still linear
equation contains the
classical spherical model as the limit case of paramagnetic high temperature state.
It is suggested that, since the restriction is now over finite volumes  in the modified model,
once the correlation length becomes greater than this typical volumes size the system must
fall into the Ising universality class.

\begin{acknowledgments}
We acknowledge D. G. Barci for useful comments in the early stages of this work.
\end{acknowledgments}

\appendix

\section{Geometrical approach to some spatial correlations.}
\label{ap1}

Firstly we are intended to demonstrate, by means of geometric
considerations, that spatial correlation C(R) in an ensemble of
perfect striped configurations has an algebraic decaying term of the
form $1/R^{(d-1)/2}$, identical to that of expression (\ref{crr}). One can perform
such calculation by simply compute the spatial average
\begin{equation}
C(R)=\langle\phi(\vec{x})\phi(\vec{x}+\vec{R})\rangle
\end{equation}
with the order parameter taken as
\begin{equation}
\phi(\vec{x})=\phi_0\ \mathrm{cos}(\vec{k}_0\cdot\vec{x}+\Phi)
\end{equation}
where $\Phi\in[0, 2\pi]$ is included to represent the arbitrary
relative position between the modulated structure and the axes origin.

\subsubsection*{$d=1$}

In one dimension the modulated structure can be written
\begin{equation}
\phi(x)=\phi_0\ \mathrm{cos}(k_0 x+\Phi)
\end{equation}
and the integration is simply over $\Phi$ and $x$. Here, $\Phi$ is the random variable containing all the relative displacements (phase)
of the modulated structure with respect to the position $x$.
The probability of finding a value between $\Phi$ and $\Phi+d\Phi$ is equal to $d\phi/2\pi$. On the other hand, the spatial 
average is performed multiplying the spatial integral with limits $\pm L$ 
by a normalization factor $1/2L$. So, we have to carry out the calculation
\begin{eqnarray}
\nonumber
C(R)&=&\lim_{L\to\infty} \int_0^{2\pi}\frac{d\Phi}{2\pi} \int_{-L}^{L}\frac{dx}{2 L}\ \\\nonumber
&\times&\  \phi_0\ \cos(k_0 x+\Phi+k_0 R)\\
&\times&\phi_0\ \cos(k_0 x+\Phi)
\end{eqnarray}
that leads to a perfect sinusoidal result
\begin{equation}
C(R)=\frac{\phi_0^2}{2}\cos(k_0 R)
\label{crd1}
\end{equation}

\subsubsection*{$d=2$}

Here we have to consider also the multiple angular orientations of the modulated structures $\vec{k}_0$ with respect to the
axes origin. Now we have
\begin{equation}
\phi(\vec{x})=\phi_0\ \mathrm{cos}(\vec{k}_0\cdot\vec{x}+\Phi)
\end{equation}
Using polar coordinates we may write $\vec{k}_0=(k_0,\alpha)$ and $\vec{x}=(r,\theta)$. 
Now the random variables are the phase $\Phi$ and the plane orientation of the structure $\alpha$. In this way,
the probability of having a stripes configuration with values around certain $\Phi$ and $\alpha$ is
\begin{equation}
P(\Phi,\alpha)d\Phi\ d\alpha=\frac{d\Phi}{2\pi}\frac{d\alpha}{2\pi}
\end{equation} 
And we have to integrate also the spatial average in polar coordinates. So, the calculation is
\begin{eqnarray}
\nonumber
C(R)&=&\lim_{L\to\infty} \int_0^{2\pi}\frac{d\Phi}{2\pi} \int_0^{2\pi}\frac{d\alpha}{2\pi} \int_0^{2\pi}\int_{0}^{L}\frac{r\ dr\ d\theta}{\pi L^2}\\\nonumber
&\times&\  \phi_0\ \cos(\vec{k}_0\cdot\vec{x}+\Phi+\vec{k}_0\cdot\vec{R})\\
&\times&\phi_0\ \cos(\vec{k}_0\cdot\vec{x}+\Phi)
\end{eqnarray}
leading to
\begin{equation}
C(R)=\frac{\phi_0^2}{2}\mathrm{J}_0(k_0 R)=\frac{\phi_0^2}{2}\frac{\cos(k_0 R-\pi/4)}{(k_0 R)^{1/2}}
\label{crd2}
\end{equation}

\subsubsection*{$d=3$}

Following the same analysis one starts now from an expression of $\phi$
\begin{equation}
\phi(\vec{x})=\phi_0\ \mathrm{cos}(\vec{k}_0\cdot\vec{x}+\Phi)
\end{equation}
in which $\vec{k}_0$ and $\vec{x}$ are three-dimensional vectors thus
representing alternating oriented laminae well-known in literature. 
With a reasoning very close to that of $d=2$, we use
spherical coordinates and write vector as $\vec{k}_0=(k_0,\alpha,\psi)$ and $\vec{x}=(r,\theta,\varphi)$.
The multiple orientations are given now by $\alpha$ and $\psi$ and the stripes are characterized by fixing
$\Phi$, $\alpha$ and $\psi$ in such a way that
\begin{equation}  
P(\Phi,\alpha,\psi)d\Phi\ d\alpha\ d\psi=\frac{d\Phi}{2\pi}\frac{d\alpha}{2\pi}\frac{\sin(\psi)d\psi}{2}
\end{equation} 
and taken the proper normalization for the spatial average in spherical coordinates 
\begin{eqnarray}
\nonumber
C(R)&=&\lim_{L\to\infty} \int_0^{2\pi}\frac{d\Phi}{2\pi} \int_0^{\pi}\int_0^{2\pi}
\frac{\sin(\psi)\ d\psi\ d\alpha}{4\ \pi} \\\nonumber
&\times&\int_0^{\pi}\int_0^{2\pi}\int_{0}^{L}\frac{r^2\ \sin(\theta)\ dr\ d\theta\ d\varphi}{\frac{4}{3}\pi L^3}\\\nonumber
&\times&\  \phi_0\ \cos(\vec{k}_0\cdot\vec{x}+\Phi+\vec{k}_0\cdot\vec{R})\\
&\times&\phi_0\ \cos(\vec{k}_0\cdot\vec{x}+\Phi)
\end{eqnarray}
This leads to 
\begin{equation}
C(R)=\frac{\phi_0^2}{2}\frac{\sin(k_0 R)}{k_0 R}=\frac{\phi_0^2}{2}\frac{\cos(k_0 R-\pi/2)}{(k_0 R)}
\label{crd3}
\end{equation}

Finally, from expressions (\ref{crd1}), (\ref{crd2}) and (\ref{crd3}) one can write, in the approximation $k_0 R \gg 1$, the
following form of $C(R)$ as function of the space dimension.
\begin{equation}
C(R)=\frac{\phi_0^2}{2}\frac{\cos[k_0 R-(d-1)\frac{\pi}{4}]}{(k_0 R)^{\frac{d-1}{2}}}
\end{equation}
and this is exactly the same functionality appearing in
(\ref{crr}) for spatial correlations, obtained in a completely different
way.

\subsection{Glassy spatial correlation}

Lets analyze now the spatial correlation in a configuration of mosaics of randomly oriented modulated structures. 
In our approach we divide the space in squared $d$-dimensional volumes of equal linear size $L$ inside which the local parameter 
is in a perfect striped pattern. The orientation of the modulated structures changes randomly from one area to
the other. Under this condition one may write
\begin{equation}
\phi(\vec{x})=\phi_0\ \mathrm{cos}(\vec{k}_0[\vec{x}]\cdot\vec{x}+\Phi[\vec{x}]) 
\end{equation}
where it has been explicitly pointed the dependence of $\vec{k}_0$ and $\Phi$ with $\vec{x}$ through squared brackets. This dependence is quite simple: while
the modulus $k_0$ is constant over all the space, the relative phase $\Phi$ and the rest of the $d-1$ values $\Omega_i$ of the orientational components of $\vec{k}_0$ 
are constant only inside the limits of the finite regions and varies randomly between regions. So, we have to compute the average
\begin{eqnarray}
C(R)&=&\phi_0^2\ \langle\cos(\vec{k}_0[\vec{x}]\cdot\vec{x}+\Phi[\vec{x}])\times \\ \nonumber
&\times&\cos(\vec{k}_0[\vec{x}+\vec{R}]\cdot(\vec{x}+\vec{R})+\Phi[\vec{x}+\vec{R}])\rangle
\end{eqnarray}
that is
\begin{eqnarray}
\nonumber
C(R)&\propto&\phi_0^2\ \int d^dx\int d^{(d-1)}\Omega\ 
\cos(\vec{k}_0[\vec{x}]\cdot\vec{x}+\Phi[\vec{x}])\times \\ 
&\times&\cos(\vec{k}_0[\vec{x}+\vec{R}]\cdot(\vec{x}+\vec{R})+\Phi[\vec{x}+\vec{R}])
\end{eqnarray}
One may note that, for $R>L$, that is, at distances higher that the typical domain size, the above expression becomes zero.
This result is consistent with the exponential decay (\ref{crfinita}) in which the non-zero value of spatial correlations establishes up to
certain length. For distances higher than this length the correlation suddenly vanishes.

\bibliography{paper_esferico_v1.0}

\end{document}